\documentclass[aps,prd,showpacs,twocolumn,preprintnumbers,amsmath,amssymb,epsf,nofootinbib,tightenlines]{revtex4}

\usepackage{graphicx}
\usepackage{bm}

\def\beqn{\begin{eqnarray}}
\def\eeqn{\end{eqnarray}}
\def\barr{\begin{array}}
\def\earr{\end{array}}
\def\btab{\begin{tabular}}
\def\etab{\end{tabular}}
\def\bite{\begin{itemize}}
\def\eite{\end{itemize}}
\def\bcen{\begin{center}}
\def\ecen{\end{center}}

\def\eq{\begin{equation}}
\def\ee{\end{equation}}
\def\nn{\nonumber}

\begin{document}

\title{Nuclear structure contribution to the Lamb shift in muonic deuterium}
\author{Carl E. Carlson}
\affiliation{College of William and Mary, 
         Physics Department,
         Williamsburg, Virginia 23187, USA}
\author{Mikhail Gorchtein}
\email{gorshtey@kph.uni-mainz.de}
\author{Marc Vanderhaeghen}
\affiliation{Institut f\"ur Kernphysik, Johannes Gutenberg-Universit\"at,   Mainz, Germany}
\affiliation{PRISMA Cluster of Excellence,  Johannes Gutenberg-Universit\"at, Mainz, Germany}
\begin{abstract}
We consider the two-photon exchange contribution to the $2P-2S$ Lamb shift in
muonic deuterium in the framework of forward dispersion relations. The
dispersion integrals are evaluated using experimental data on elastic
deuteron form factors and inelastic electron-deuteron scattering, both
in the quasielastic and hadronic range. The subtraction constant that
is required to ensure convergence of the dispersion relation for the
forward Compton amplitude $T_1(\nu,Q^2)$ is related to the deuteron
magnetic polarizability $\beta(Q^2)$. Based on phenomenological information, 
we obtain for the Lamb shift $\Delta
E_{2P-2S}=2.01\pm0.74$ meV. The main source of the
uncertainty of the dispersion analysis is due to lack of quasielastic data 
at low energies and forward angles. We show that a targeted
measurement of the deuteron electrodesintegration in the kinematics of
upcoming experiments A1 and MESA at Mainz can help quenching this
uncertainty significantly. 
\end{abstract}
\date{\today}
\maketitle


\section{Introduction}


The proton radius puzzle---that the proton radius obtained from the Lamb shift in muonic hydrogen~\cite{pohlmuH, antogninimuH} is different from what should be the same radius obtained from data involving electrons~\cite{Mohr:2012tt,bernauer}---has attracted much attention in recent years.  The explanation of the problem is not known so far.  
There are proposals of new dedicated scattering experiments with electrons \cite{gasparyan,ISR_A1}
and muons \cite{Gilman:2013eiv}. On the theory side, the discrepancy was addressed in terms of effective non-relativistic QED interactions \cite{Hill:2011wy}, dispersion relations ~\cite{Lorenz:2012tm}, exotically large hadronic effects~\cite{Miller:2012ne}, or of new physics affecting the muon and electron differently~\cite{TuckerSmith:2010ra,Batell:2011qq,Carlson:2012pc}.\\

Further information can come from measuring the deuteron radius using the Lamb shift in muonic deuterium.  The deuteron radius from electron based experiments is already known to good accuracy.  The best results come from using the isotope shift, that is, measuring the $1S$-$2S$ splittings in electron-proton ($e$-$H$) and electron-deuteron ($e$-$D$) hydrogen and finding from residual corrections that
\begin{equation}
r^2_E(d) - r^2_E(p) = 3.82007(65) {\rm\ fm^2}	\,,
\end{equation}
as quoted in~\cite{Parthey:2010,huber}, where the $r_E(p,d)$ are charge radii.  The isotope shift number is so accurate that the uncertainty in the deuteron radius-squared becomes in practice the same as for the proton, and using the CODATA 2010 value for the proton radius one finds~\cite{Mohr:2012tt},
\begin{equation}
r_E(d) = 2.1424 (21) {\rm\ fm}	\,.
\end{equation}
The uncertainty is $0.1\%$.  In contrast, the current best direct electron-deuteron scattering results yield $r_E(d)=2.128(11)$ fm~\cite{Sick:1996}, or $0.5\%$ uncertainty.  Planned experiments are expected to reduce this uncertainty~\cite{modkg}.

To obtain the charge radius from the Lamb shift requires not only accurate data but also accurate calculation of all corrections that are not hadronic size corrections.  Of these, the two-photon correction, which includes the relativistically correct polarizability correction, has drawn continued attention.  A deuteron is easily distorted compared to a single proton, and we shall see that the polarizability correction for the $\mu$-$D$ system is about two orders of magnitude larger than for $\mu$-$H$.  The requirement that the $\mu$-$D$ polarizability correction be safely smaller than the radius-related energy shift can become quite severe.

Of course, without knowing the underlying reason for the proton radius discrepancy, we cannot with certainty predict what the deuteron discrepancy will be.  However, we will give the anticipated energy discrepancy in one scenario, and thereby obtain a working number with the expectation that other scenarios would give results similar within a factor of a few.  As a reminder, the main energy shift due to finite hadron (or nuclear) size is
\begin{equation}
\Delta E_{\rm finite\ size} = \frac{2\pi Z\alpha}{3} \frac{(m_r Z\alpha)^3}{n^3 \pi} r^2_E(h) \,,
\end{equation}
for a hydrogen-like atom in the $nS$ state, where $m_r$ is the reduced mass.  Experiment shows about a $320\,\mu$eV energy discrepancy for the $\mu$-$H$ $2S$ state, compared to expectations based on the CODATA 2010~\cite{Mohr:2012tt} electron based proton radius.

In a scenario where the $\mu$-$H$ energy discrepancy is not actually due to a proton size change but rather to the exchange of a new particle that has a special coupling to the muon and {\it e. g.} a dark photon coupling (\textit{i.e.,} a squelched electromagnetic coupling) to other particles, the $\mu$-$D$ energy would be the same as in the $\mu$-$H$ except for the reduced-mass-cubed factor.  In this case the anticipated $\mu$-$D$ energy discrepancy is 
\begin{equation}
\Delta E_{\rm discrepant}(\mu{\rm -}D) \approx 380 \ \mu{\rm{eV}}.
\end{equation}
Hence the two-photon corrections should be calculated within $100\ \mu$eV or less.  As the two-photon corrections are of order $2$ meV, this requires a $5\%$ or better accuracy.

There are several currently available polarizability calculations~\cite{friar,pachuckiD,rosenfelder,FriarPayne,Friar:2013rha,Ji:2013oba}.  Ref.~\cite{pachuckiD} includes the elastic contributions and quotes an uncertainty limit well within the requirement. A more recent calculation of Ref.~\cite{Ji:2013oba} follows the same lines but includes further relativistic corrections. In addition, much of that calculation is supported by a calculation ~\cite{Friar:2013rha} that uses the zero-range approximation which indicates that the bulk of the result is relatively model independent and can be obtained non-relativistically.  However, there remain significant energy shifts that are obtained non-relativistically using a potential model.  One would like a calculation by an alternative method to verify the results.

We here explore a fully relativistic dispersive calculation of the two-photon corrections to $\mu$-$D$. In this type of calculation, the energy shift is obtained from the real part of an amplitude whose imaginary part is related without approximation to physical elastic and inelastic $e$-$D$ scattering.  To the extent that the data is accurate and sufficient, the result follows just from the data and the calculation is model independent.  The calculation resembles the generally accepted work for the $\mu$-$H$ system~\cite{Pachucki:1996zza,Martynenko:2005rc,Carlson:2011zd,Birse:2012eb,fesr}.  However, for the proton case the corrections are much smaller, and the proton data in the relevant kinematic regions is itself very good. Furthermore, good analytic fits, useful for doing integrals, are already available in the literature~\cite{bostedchristy}.

For the deuteron we need elastic and inelastic data over a wide range of energies, including the low energy region, where the data is sparse.  There are good analytic fits~\cite{bostedchristy} to the deuteron data above the pion production threshold, but not in the lower energy quasi-elastic region.  Part of our effort is devoted to providing such fits.  The inelastic data is represented in terms of structure functions or response functions. The overall polarizability effect is sensitive to the structure of the response functions at low energy and low virtual photon masses, which is where the data is sparse.  A consequence of this, phrased in terms of fitting procedures, is that small changes in some parameters have big effects on the near threshold behavior but small effect on the quality of the fits to the available data.  This leads to a larger than desired uncertainty in the results for the two-photon corrections, when using this method.

Future low momentum transfer deuteron breakup data, possibly obtained as background data to elastic scattering experiments~\cite{modkg}, can bring about a decisive reduction in the uncertainty limits of the polarizability calculation.  We discuss this below with some examples.

Our presentation starts with the description of the general dispersive formalism for obtaining energy shifts from elastic and inelastic scattering data and subtraction terms in Section II. Evaluation of the elastic and inelastic dispersion integrals is discussed in Section III where we as well provide details of our global fit to the available quasi-elastic deuteron data. In Section IV we present our results, discuss the uncertainty limits, and address possible help from upcoming experiments. Section V is dedicated to the application of the dispersive, data-based, model-independent polarizability calculation described here to the $e$-$H$ and $e$-$D$ systems and its relevance to the isotope shift measurements. Section VI closes the article with a short summary of our findings.

\section{The Basic Formalism}

\begin{figure}[h]
\includegraphics[height=2cm]{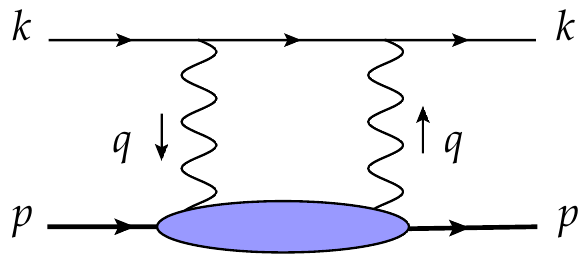}
\caption{(Color online) Two-photon exchange diagram for the ${\cal O}(\alpha^5)$
  correction to the Lamb shift.}
\label{tpediag}
\end{figure}

The diagram that contains the nuclear and hadronic structure-dependent
${\cal O}(\alpha^5)$ correction to the Lamb shift is shown in
Fig. \ref{tpediag}. 

 The lower part of the diagram, the blob containing the
nuclear and hadronic structure dependence is encoded in the forward
virtual Compton tensor,
\beqn
&&T^{\mu\nu}=\frac{i}{8\pi M_d}\int d^4xe^{iqx}\langle
p|T\,j^\mu(x)j^\nu(0)|p\rangle\\
&&=\left(-g^{\mu\nu}+\frac{q^\mu
    q^\nu}{q^2}\right)T_1(\nu,Q^2)
+\frac{\hat p^\mu\hat p^\nu}{M_d^2} T_2(\nu,Q^2),\nn
\eeqn
where $\hat p^\mu=p^{\mu}-\frac{p\cdot q}{q^2}q^\mu$, $Q^2=-q^2$, 
$\nu=(p\cdot q)/M_d$ and $M_d$ is the deuteron
mass. 
Following \cite{Carlson:2011zd}, we can write the contribution of the
two-photon exchange diagram to the $n\ell$ energy level as 
\beqn
&&\Delta E_{n\ell}=\frac{8\alpha^2m}{\pi}\phi^2_{n\ell}(0)\int d^4Q\\
&&\times\frac{(Q^2+2Q_0^2)T_1(iQ_0,Q^2)-(Q^2-Q_0^2)T_2(iQ_0,Q^2)}{Q^4(Q^4+4m^2Q_0^2)},\nn
\eeqn
where a Wick rotation $q_0=iQ_0$ was made, and
$\phi^2_{n\ell}(0)=\mu_r^3\alpha^3/(\pi n^3)\delta_{\ell0}$, $\mu_r=mM/(M+m)$ being the
reduced mass. $T_{1,2}(\nu,Q^2)$ are even functions of $\nu$ and
their imaginary parts are related to the spin-independent structure
functions of lepton-deuteron scattering,
\beqn
{\rm Im}T_1(\nu,Q^2)&=&\frac{1}{4M_d}F_1(\nu,Q^2)\nn\\
{\rm Im}T_2(\nu,Q^2)&=&\frac{1}{4\nu}F_2(\nu,Q^2).
\eeqn
Given the known high-energy behavior of the structure functions, the
two amplitudes obey the following form of dispersion relation, 
\beqn
{\rm Re}T_1(q_0,Q^2)&=&\bar T_1(0,Q^2)+{\rm Re}T_1^{pole}(q_0,Q^2)\\
&+&\frac{q_0^2}{2\pi M_d}\int\limits_{\nu_{thr}}^\infty\frac{d\nu F_1(\nu,Q^2)}{\nu(\nu^2-q_0^2)}\nn\\
{\rm Re}T_2(q_0,Q^2)&=&{\rm Re}T_2^{pole}(q_0,Q^2) 
+\frac{1}{2\pi}\int\limits_{\nu_{thr}}^\infty\frac{d\nu F_2(\nu,Q^2)}{\nu^2-q_0^2},\nn
\eeqn
where for $T_1$ the subtraction at $q_0=0$ was performed with $\bar T_1(0,Q^2)$ the respective subtraction function.
Above, we explicitly extracted the contribution of the ground state
leading to a pole, $T_{1,2}^{pole}$. This contribution is defined in
terms of the deuteron's electromagnetic vertex
\beqn
&&\langle d(p')|J^\mu(q)|d(p)\rangle=G_2(Q^2)[{\xi'^*}^\mu(\xi q)-\xi^\mu(\xi'^* q)]\nn\\
&&-\left[G_1(Q^2)(\xi'^*\xi)-G_3(Q^2)\frac{(\xi'^* q)(\xi
    q)}{2M_d^2}\right](p+p')^\mu\!\!\!,\;\;\;\;
\eeqn
where $\xi^\mu({\xi'^*}^\mu)$ denote the polarization vector of the
initial (final) deuteron with momenta $p(p')$, respectively, and
$Q^2=-q^2$ stands for the four-momentum transfer. 
The
form factors $G_{1,2,3}$ are related to the charge, magnetic and
quadrupole deuteron form factors as
\beqn
G_M&=&G_2,\nn\\
G_C&=&G_1+\frac{2}{3}\tau_dG_Q,\nn\\
G_Q&=&G_1-G_2+(1+\tau_d)G_3,
\eeqn
and $\tau_d=Q^2/(4M_d^2)$. 
The elastic contribution to the structure functions reads
\beqn
F_1^{el}&=&\frac{1}{3}(1+\tau_d)G_M^2\delta(1-x_d),\\
F_2^{el}&=&\left[G_C^2+\frac{2}{3}\tau_dG_M^2+\frac{8}{9}\tau_d^2G_Q^2\right]\delta(1-x_d),\nn
\eeqn
with the Bjorken variable $x_d=Q^2/(2M_d\nu)$.

Correspondingly, we distinguish three contributions, 
$\Delta E_{n0}=\Delta E_{n0}^{subt}+\Delta E_{n0}^{el}+\Delta E_{n0}^{inel}$ where
\beqn
\Delta
E_{n0}^{subt}&=&\frac{4\pi\alpha^2}{m}\phi^2_{n0}(0)\int_0^\infty\frac{dQ^2}{Q^2}\frac{\gamma_1(\tau_l)}{\sqrt\tau_l}\bar
T_1(0,Q^2),\label{subtraction}
\eeqn
\beqn
\Delta E^{el}_{n0}&=&\frac{m\alpha^2}{M_d(M_d^2-m^2)}\phi^2_{n0}(0)
\int_0^\infty\frac{dQ^2}{Q^2}\label{elastic}\\
&\times&\left\{
\frac{2}{3}G_M^2
(1+\tau_d)\left(\frac{\gamma_1(\tau_d)}{\sqrt\tau_d}
-\frac{\gamma_1(\tau_l)}{\sqrt\tau_l}\right)
\right.\nn\\
&&-\left.
\left(\frac{\gamma_2(\tau_d)}{\sqrt\tau_d}-\frac{\gamma_2(\tau_l)}{\sqrt\tau_l}\right)
\left[\frac{G_C^2}{\tau_d}+\frac{2}{3}G_M^2+\frac{8}{9}\tau_d G_Q^2\right]
\right\}\nn
\eeqn
\beqn
\Delta E^{inel}_{n0}&=&-\frac{2\alpha^2}{M_dm}\phi^2_{n0}(0)
\int_0^\infty\frac{dQ^2}{Q^2}\int_{\nu_{thr}}^\infty\frac{d\nu}{\nu}\label{inelastic}\\
&\times&\left[
\tilde\gamma_1(\tau,\tau_l)F_1(\nu,Q^2)
+\frac{M_d\nu}{Q^2}\tilde\gamma_2(\tau,\tau_l)F_2(\nu,Q^2)
\right].\nn
\eeqn
Above, we denote $\tau=\nu^2/Q^2$, $\tau_l=Q^2/(4m^2)$, and the 
auxiliary functions are given by
\beqn
\gamma_1(\tau)&=&(1-2\tau)\sqrt{1+\tau}+2\tau^{3/2}\nn\\
\gamma_2(\tau)&=&(1+\tau)^{3/2}-\tau^{3/2}-\frac{3}{2}\sqrt \tau\nn\\
\tilde\gamma_1(\tau,\tau_l)&=&\frac{\sqrt \tau\gamma_1(\tau)-\sqrt \tau_l\gamma_1(\tau_l)}{\tau-\tau_l}\nn\\
\tilde\gamma_2(\tau,\tau_l)&=&\frac{1}{\tau-\tau_l}
\left[\frac{\gamma_2(\tau_l)}{\sqrt \tau_l}-\frac{\gamma_2(\tau)}{\sqrt \tau}\right].
\eeqn


\section{Evaluation and Data Fits}


\subsection{Elastic contribution}
We start with the elastic contribution. It can be noted that the
integral in Eq. (\ref{elastic}) is IR divergent due to an exchange of
soft Coulomb photons. Such contributions, however, were already taken
into account within the non-relativistic calculations on a pointlike
deuteron. Furthermore, the finite size effects were 
accounted for, as well, and have to be subtracted from the full result of
Eq. (\ref{elastic}) to avoid double-counting. 
This subtraction leads to
\beqn
&&\Delta \bar E^{el}_{n0}=\frac{m\alpha^2}{M_d(M_d^2-m^2)}\phi^2_{n0}(0)
\int_0^\infty\frac{dQ^2}{Q^2}\label{elastic_subt}\\
&&\times\left\{
\frac{2}{3}G_M^2
(1+\tau_d)\left(\frac{\gamma_1(\tau_d)}{\sqrt\tau_d}
-\frac{\gamma_1(\tau_l)}{\sqrt\tau_l}\right)
\right.\nn\\
&&-
\left(\frac{\gamma_2(\tau_d)}{\sqrt\tau_d}-\frac{\gamma_2(\tau_l)}{\sqrt\tau_l}\right)
\left[\frac{G_C^2-1}{\tau_d}+\frac{2}{3}G_M^2+\frac{8}{9}\tau_d G_Q^2\right]\nn\\
&&+16M_d^2\frac{M_d-m}{Q}G_C'(0)
\Big\}.\nn
\eeqn
\indent
We evaluate Eq. (\ref{elastic_subt}) with the most recent deuteron form
factors' parametrization from \cite{abbott}. We use the
parametrization I and II of that Ref. to estimate the uncertainty, and
list the result with the uncertainty in Table \ref{tab2}.\\

The inelastic contributions contain two parts, 
\beqn
\Delta E^{inel}_{n0}=\Delta E^{QE}_{n0}+\Delta E^{hadr}_{n0},
\eeqn 
the quasielastic nucleon
knock-out ({\it QE}) and hadronic excitation spectrum ({\it hadr}) that we will treat separately. \\

\subsection{Hadronic contribution}
The part of the deuteron excitation spectrum above the
pion production threshold can be dealt with very similarly as it was
done in Ref. \cite{Carlson:2011zd} for the proton case. We use the modern
deuteron virtual photoabsorption data that were parametrized in terms
of resonances plus non-resonant background by Bosted and Christy in
\cite{bostedchristy}. Since the integration over the energy extend
beyond the validity of the fit of Ref. \cite{bostedchristy}, we
supplement the correct high-energy behavior by adopting a
Regge-behaved background. The Regge fit to world data on the deuteron
total photoabsorption cross section was done in \cite{fesr}. We extend
this description to virtual photoabsorption by supplementing a
$Q^2$-dependence from generalized VDM, e.g. \cite{alwall}, that
provides good description of virtual photoabsorption data at
$Q^2\lesssim3$ GeV$^2$. 
The result for $\Delta E^{hadr}$ is reported in Table \ref{tab2}. \\

\subsection{Quasielastic contribution}
In the literature, there exist non-relativistic calculations of the Lamb shift in
muonic deuterium with potential models or in zero-range approximations
\cite{friar,pachuckiD,rosenfelder,FriarPayne,Friar:2013rha,Ji:2013oba}. In this work we opt for a
phenomenological, data-driven approach in the spirit of
Ref. \cite{Carlson:2011zd} where real and
virtual photoabsorption data on the proton 
were utilized to constrain the Lamb shift in the muonic hydrogen. For
this purpose we need to fit the quasi-elastic data over the whole
kinematical range with an appropriate function of $\nu,\,Q^2$. 

We start with the plane wave Born approximation (PWBA) that allows
to relate the deuteron structure functions in the quasi-elastic
kinematics to the elastic nucleon structure functions. In doing this,
we Fermi-smear the nucleon virtual Compton tensor, rather than just
the structure functions, the result reads
\beqn
F_1^{PWBA}&=&
\frac{Q^2}{4|\vec q|}G_M^2 S(\nu,Q^2)+\frac{1}{2|\vec q|}\frac{G_E^2+\tau
    G_M^2}{1+\tau}S_\perp(\nu,Q^2)\nn\\
F_2^{PWBA}&=&\frac{\nu Q^2}{2M_d|\vec q|^3}\frac{G_E^2+\tau
    G_M^2}{1+\tau}\nn\\
&\times&\left[(2M+\nu)^2\frac{Q^2}{2|\vec q|^2}S(\nu,Q^2)+S_\perp(\nu,Q^2) \right].
\eeqn
Above, we defined the integrals
\beqn
S(\nu,Q^2)&=&\int\limits_{k_{min}}^{k_{max}}dk\,k(u^2(p)+w^2(p)),\nn\\
S_\perp(\nu,Q^2)&=&\int\limits_{k_{min}}^{k_{max}}dk\,k(u^2(p)+w^2(p))
k_\perp^2
\eeqn
The deuteron wave function is normalized as 
$\int k^2dk[u^2(p)+w^2(p)]=1$, $u,w(p)$ denote the
$s,d$ radial wave function of the deuteron, respectively. 
The magnitude of the three-momentum
of the bound nucleon is denoted by $k=|\vec k|$, and its component
perpendicular to the direction of the virtual photon is $k_\perp^2=k^2\sin^2\theta_k$, the
angle $\theta_k$ is defined below.   
\begin{figure}
\includegraphics[width=4.5cm]{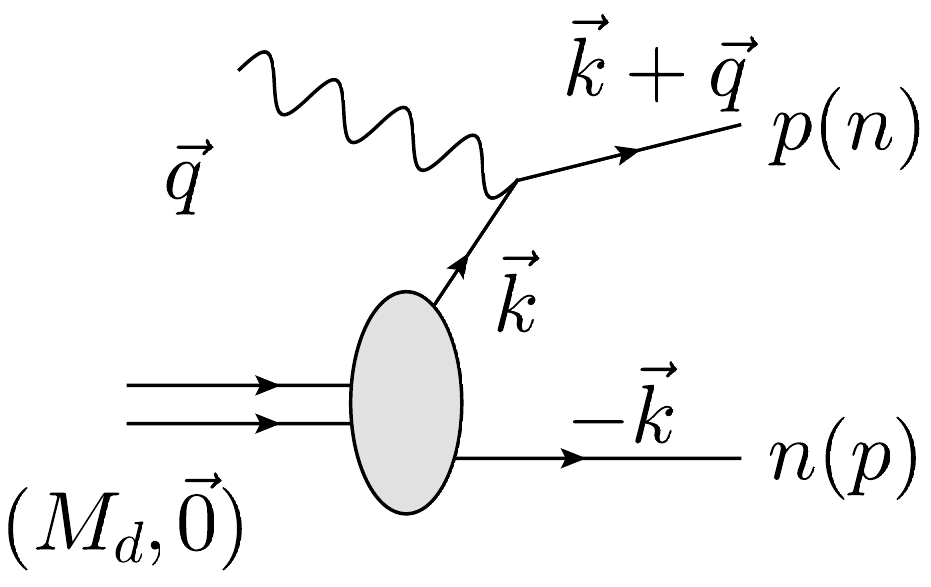}
\caption{Quasielastic scattering kinematics.}
\label{figQE}
\end{figure}
The on-shell condition for the external (knock-out) nucleons and 4-momentum conservation 
\beqn
M_d+\nu&=&\sqrt{M^2+(\vec q+\vec k)^2}+\sqrt{M^2+\vec k^2},
\eeqn
with the average nucleon mass
$M\equiv\frac{1}{2}(M_p+M_n)\approx0.938919$ GeV,
leads to a delta function for the angle between the three-momenta of
the virtual photon and the active nucleon,
\beqn
\cos\theta_k=\frac{2(M_d+\nu)\sqrt{M^2+\vec k^2}-(M_d+\nu)^2+
|\vec q|^2}{2|\vec q||\vec k|}.
\eeqn
The integral over the Fermi momentum $k$
is constrained between two finite values due to the requirement $-1\leq\cos\theta_k\leq1$,
\beqn
k_{min}&=&\left|-\frac{|\vec q|}{2}+\frac{M_d+\nu}{2}\sqrt{\frac{\nu-\nu_{min}}{M_d/2+\nu-\nu_{min}}}\right|,\nn\\
k_{max}&=&\frac{|\vec q|}{2}+\frac{M_d+\nu}{2}\sqrt{\frac{\nu-\nu_{min}}{M_d/2+\nu-\nu_{min}}},
\eeqn
with $\nu_{min}=Q^2/(2M_d)+\epsilon+{\cal O}(\epsilon^2)$ and
$\epsilon=2M-M_d\approx2.224$ MeV. 

Experimental data on deuteron electrodesintegration feature a sharp
peak just above the threshold, that is due to final state interactions
between proton and neutron after the break-up \cite{Durand:1961zz}. 
We adopt an approximate formula (Eq. (48) of that Ref.) to obtain the
following model for the transverse and longitudinal cross sections,
\beqn
\sigma_T^0&=&\sqrt\frac{\nu-\nu_{min}}{M+\nu-\nu_{min}}\,\frac{[G^p_M(Q^2)-G_M^n(Q^2)]^2}{1+M(\nu-\nu_{min})a_S^2}\\
\sigma_L^0&=&\sqrt\frac{\nu-\nu_{min}}{M+\nu-\nu_{min}}\,\frac{[G^p_E(Q^2)+G_E^n(Q^2)]^2}{1+M(\nu-\nu_{min})a_T^2},
\eeqn
with the $n-p$ singlet and triplet scattering lengths entering the FSI
part, $a_S=-23.74$ fm, $a_T=5.38$ fm, respectively. Above, we note
that the combination $\sqrt{M(\nu-\nu_{min})}=|\vec p|$ corresponds to
the three-momentum of the knocked-out nucleon. 

As a result, we obtain the following representation of the
quasielastic structure functions of the deuteron:
\beqn
F_{1,2}^{d,\,QE}=F_{1,2}^\perp+F_{1,2}^{PWBA}+F_{1,2}^{FSI},\label{model1}
\eeqn
according to the ingredients discussed above. To describe data at
arbitrary kinematics, we allow for a rescaling of each ingredient by a
function of $Q^2$ that should be obtained from the fit to the deuteron
photo- and electrodesintegration data. 
\begin{eqnarray}
F_1^\perp&=&C_\perp\frac{G_E^2+\tau
  G_M^2}{1+\tau}\frac{1}{2|\vec q|}S_\perp(\nu,Q^2),\nn\\
F_2^\perp&=& \frac{\nu Q^2}{M_d|\vec q|^2}F_1^\perp,\nn\\
F_1^{PWBA}&=&f_T^{PWBA}\!\!(Q^2)\frac{Q^2}{4|\vec q|}G_M^2 S(\nu,Q^2),\nn\\
F_2^{PWBA}&=&f_T^{PWBA}\!\!(Q^2)\frac{\nu Q^4}{M_d|\vec q|^5}\frac{G_E^2+\tau
  G_M^2}{1+\tau}\nn\\
&\times&S(\nu,Q^2) (M+\frac{\nu}{2})^2,\nn\\
F_1^{FSI}&=&M_d f_T^{FSI}(Q^2)\sigma_T^0,\nn\\
F_2^{FSI}&=&\frac{\nu Q^2}{|\vec
  q|^2}(f_T^{FSI}(Q^2)\sigma_T^0+f_L^{FSI}(Q^2)\sigma_L^0),
\label{model2}
\end{eqnarray}
where we adopted the following forms: 
\beqn
f_T^{PWBA}(Q^2)&=&\left[1- a_1 e^{-b_1 Q^2}\right],\nn\\
f_T^{FSI}(Q^2)&=&\frac{100 a_2 Q^2}{(1+b_2Q^2)^2},\nn\\ 
f_L^{FSI}(Q^2)&=&\frac{1}{{\rm MeV}}\frac{1-e^{-a_3Q^2}}{1+b_3Q^2}.\label{fit}
\eeqn 
We fitted the available data from $Q^2=0.005$ GeV$^2$ to $Q^2=3$  GeV$^2$, 
and the resulting values of the parameters are listed in Table \ref{tab1}.
\begin{table}
  \begin{tabular}{c|c|c}
\hline
$a_1$ & $b_1$(GeV$^{-2}$) & $a_2$(GeV$^{-3}$) \\ 
\hline
0.995(5) & 25.4(6) & 215(35)\\
\hline\hline
$b_2$(GeV$^{-2}$) & $a_3$(GeV$^{-3}$) & $b_3$(GeV$^{-2}$)  \\
\hline
 52(8) & 3.5(1.5) & 24.5(8.0)\\
\hline
\end{tabular}
\caption{The values of the parameters introduced in Eq. (\ref{fit}) as obtained from a fit to the deuteron QE data.}
\label{tab1}
\end{table}

\begin{figure}
\includegraphics[width=8.5cm]{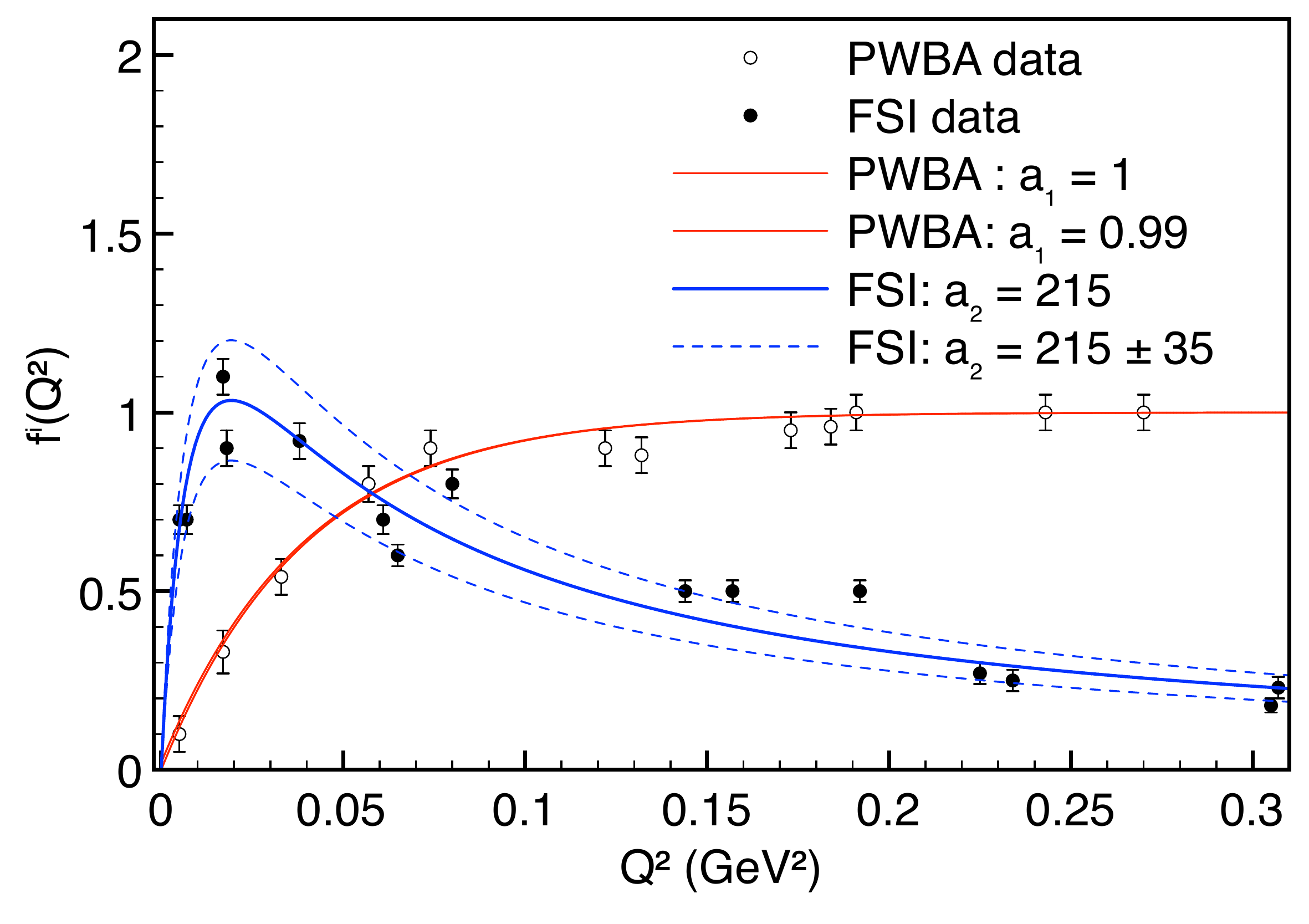}
\caption{(Color online) Scaling factors $f_T^{i}(Q^2)$ with uncertainty thereof: $i=PWBA$ (red solid lines), 
  and $i=FSI$ (blue solid line) with uncertainty thereof (thin
  blue short-dashed lines) plotted vs. QE data as function of $Q^2$. }
\label{fig1}
\end{figure}


\begin{figure}
\includegraphics[width=8.5cm]{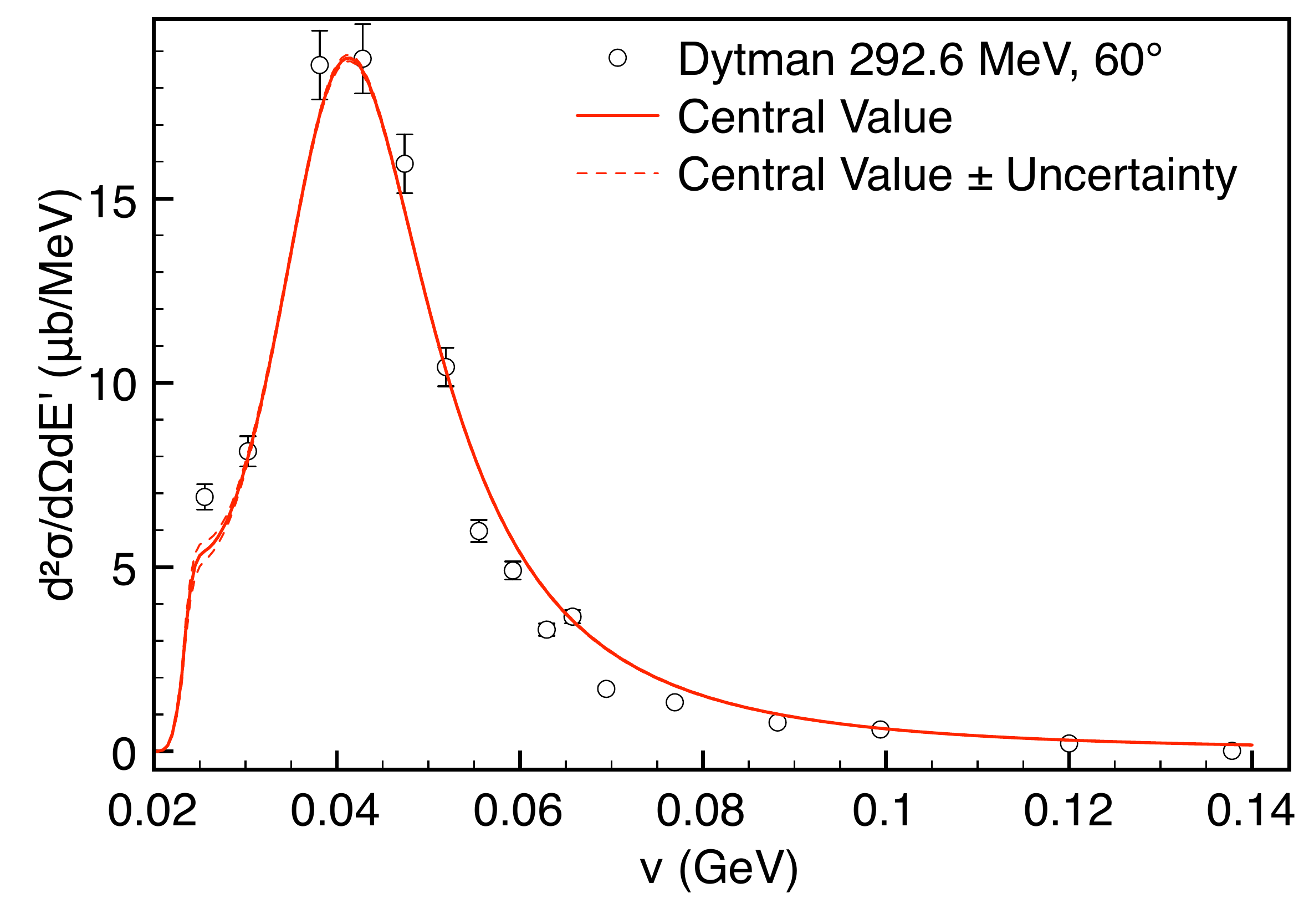}
\caption{(Color online) Rescaled PWBA model of Eqs. (\ref{model1}, \ref{model2})
vs data from Ref. \cite{dytman}.}
\label{fig2}
\end{figure}

\begin{figure}
\includegraphics[width=8.5cm]{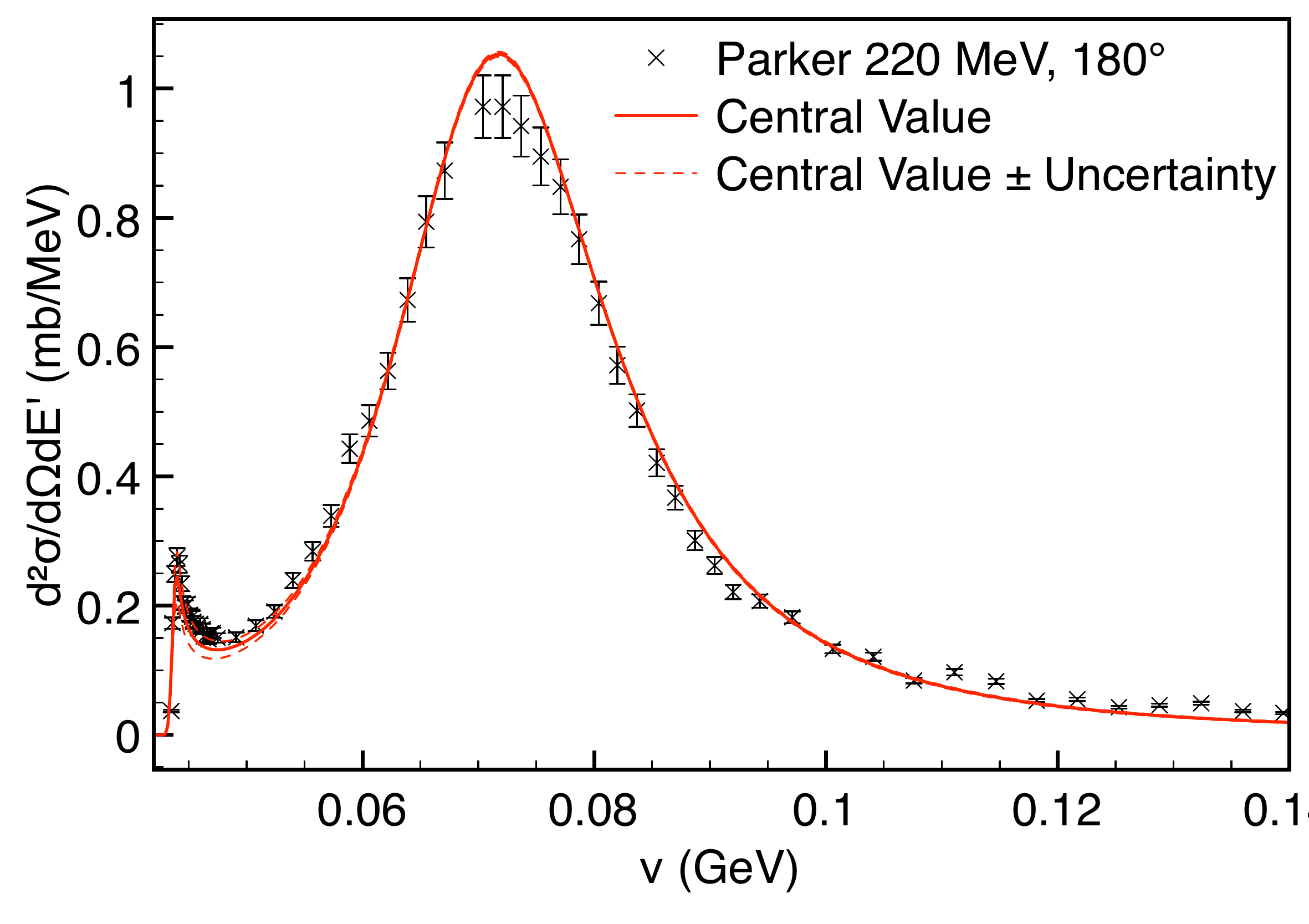}
\caption{(Color online) Same as in Fig. \ref{fig2} vs data  from Ref. \cite{parker}.}
\label{fig3}
\end{figure}

\begin{figure}
\includegraphics[width=8.5cm]{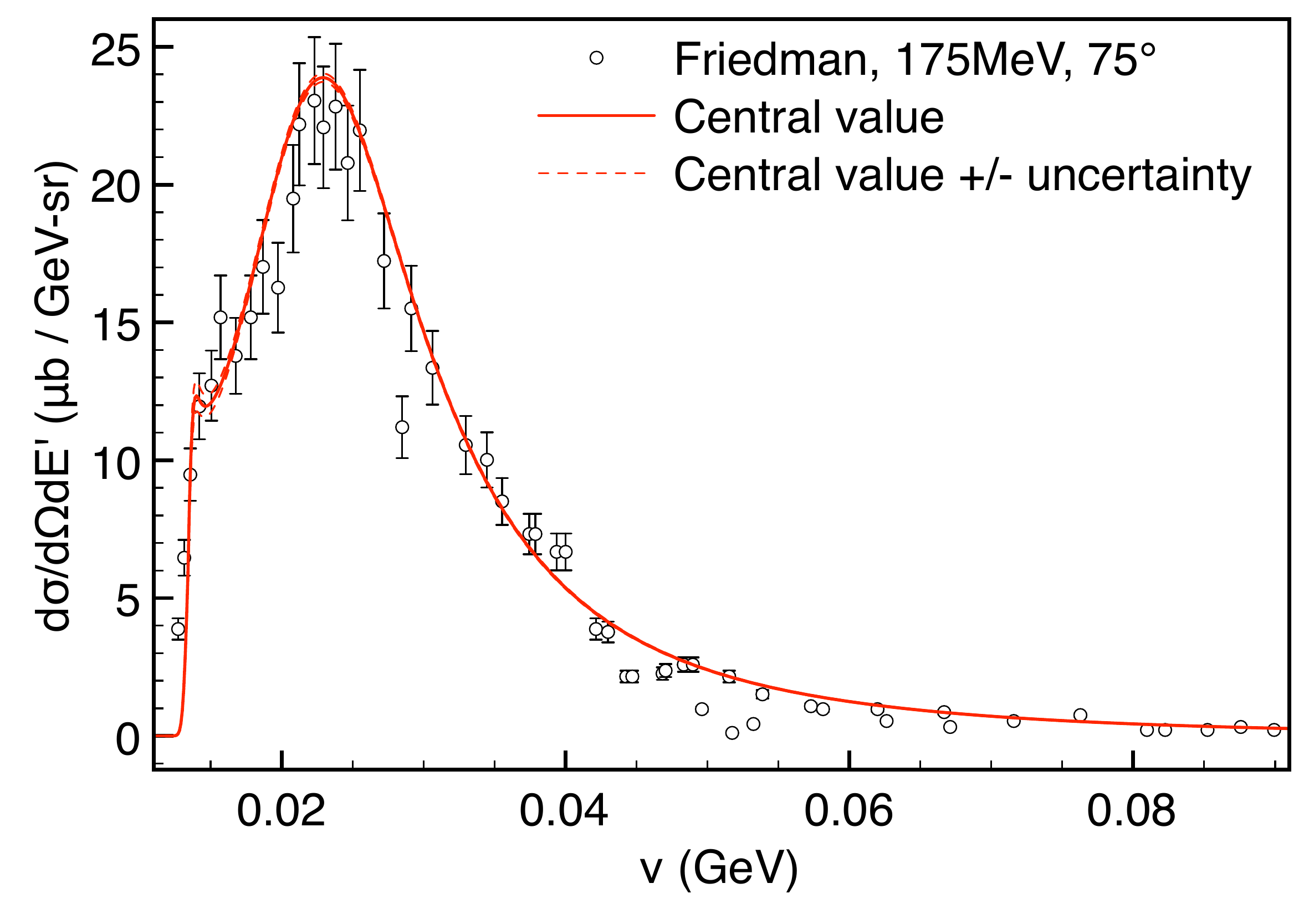}
\caption{(Color online) Same as in Fig. \ref{fig2} vs data  from Ref. \cite{friedman}.}
\label{fig4}
\end{figure}

\begin{figure}
\includegraphics[width=8.5cm]{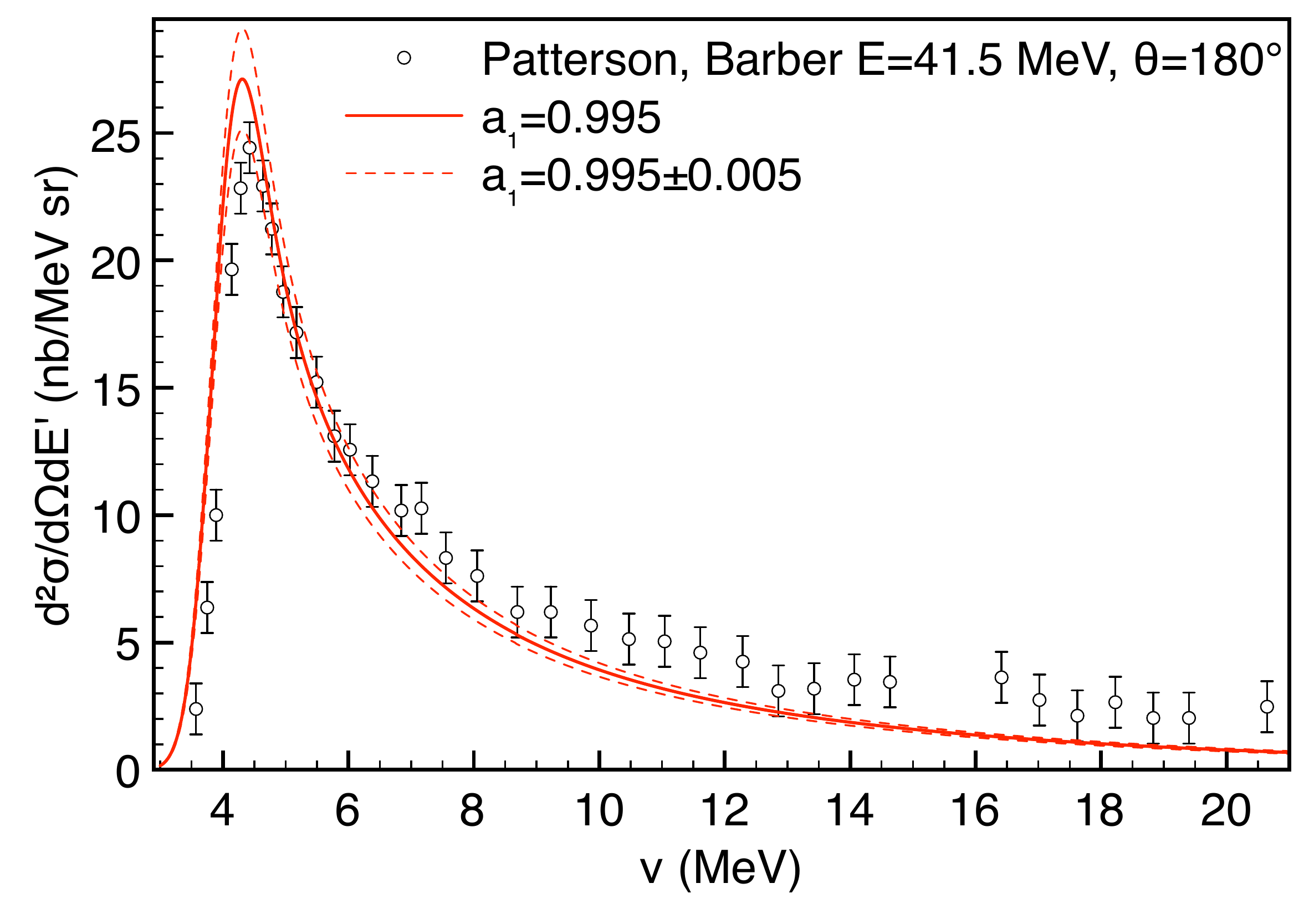}
\caption{(Color online) Same as in Fig. \ref{fig2} vs data  from Ref. \cite{patterson}.}
\label{fig5}
\end{figure}

\begin{figure}
\includegraphics[width=8.5cm]{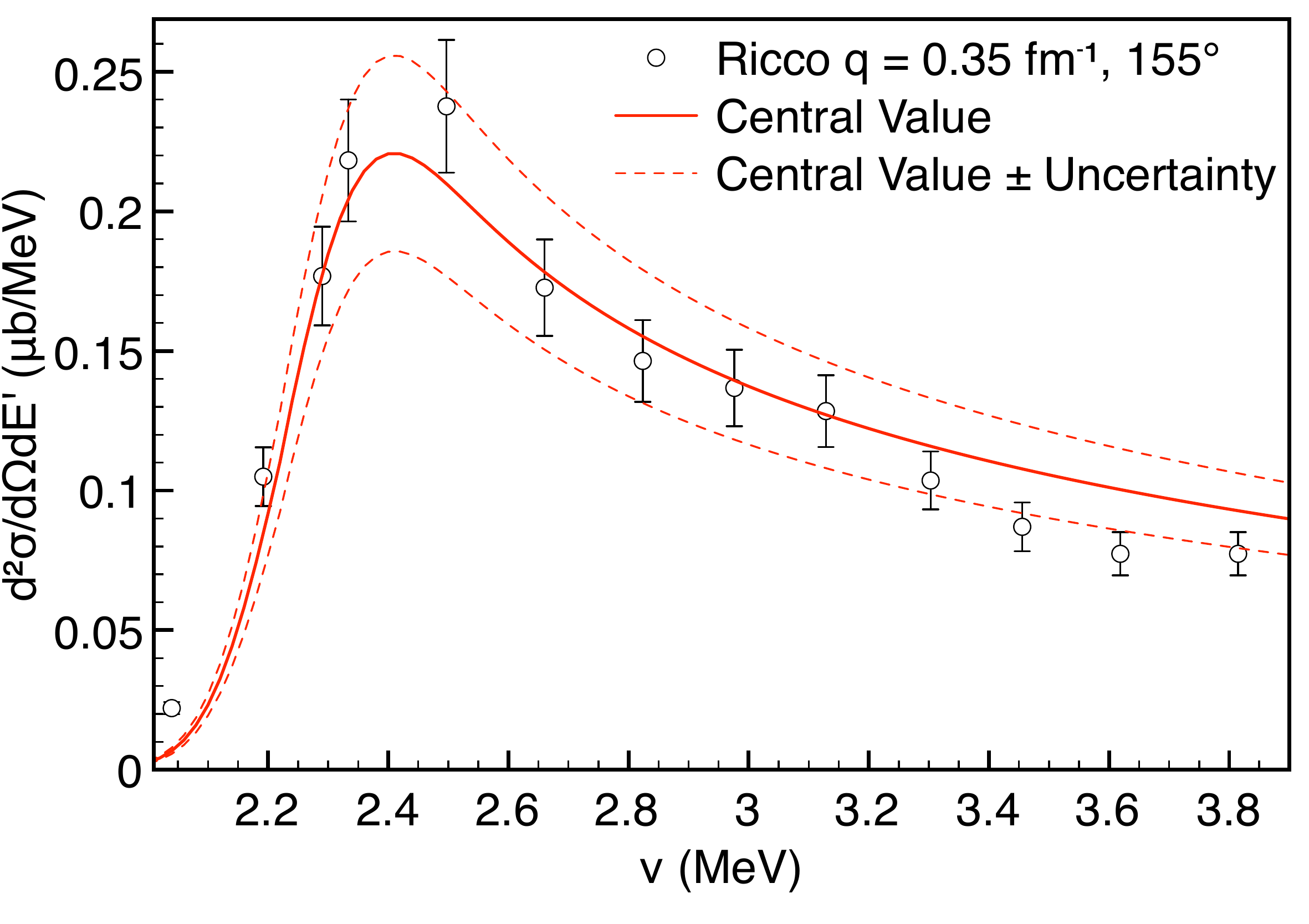}
\caption{(Color online) Same as in Fig. \ref{fig2} vs data  from Ref. \cite{ricco}.}
\label{fig6}
\end{figure}

\begin{figure}
\includegraphics[width=8.5cm]{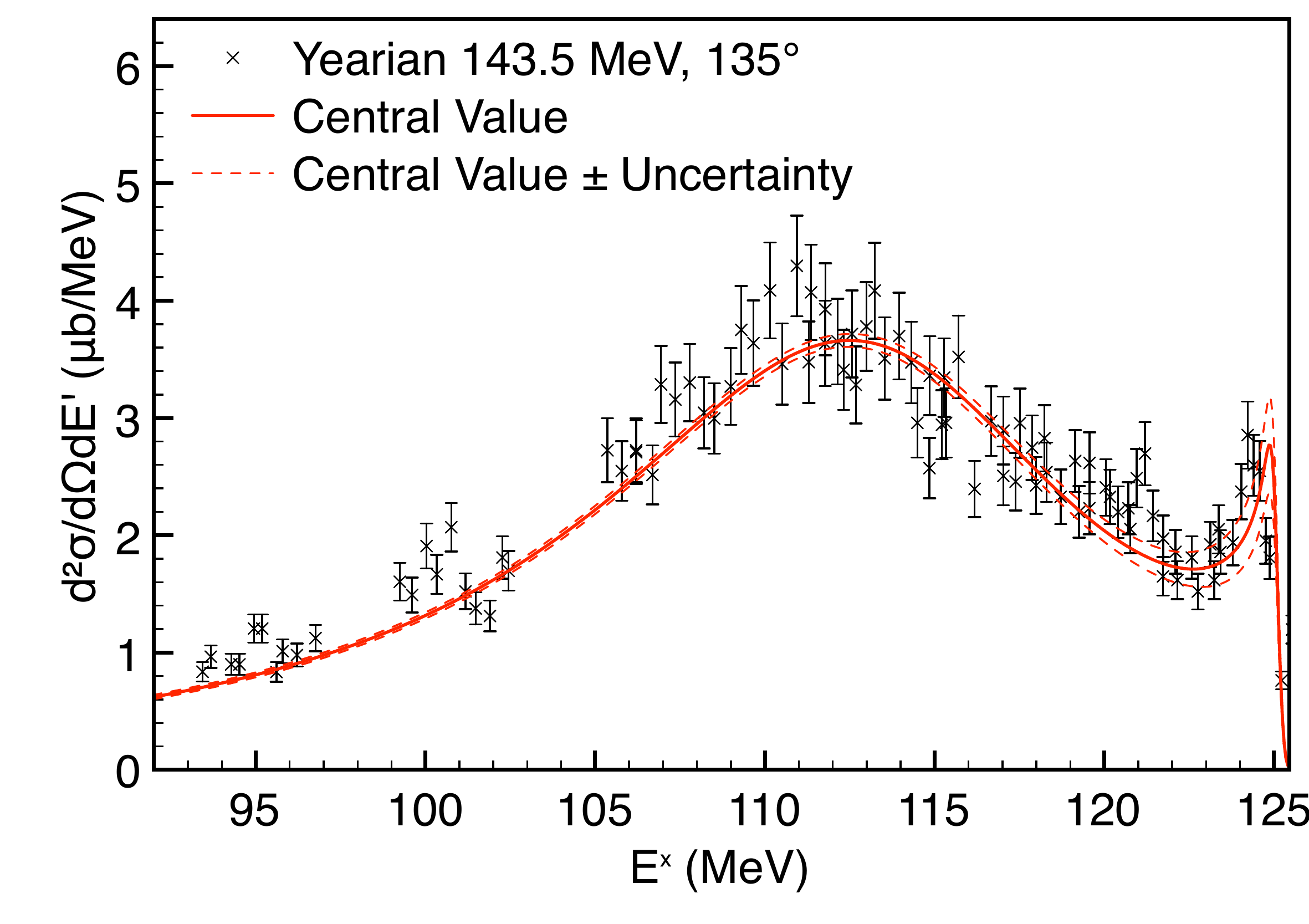}
\caption{(Color online) Same as in Fig. \ref{fig2} vs data  from Ref. \cite{yearian} as function of
  excitation energy $E^x$.}
\label{fig7}
\end{figure}

\begin{figure}
\includegraphics[width=8.5cm]{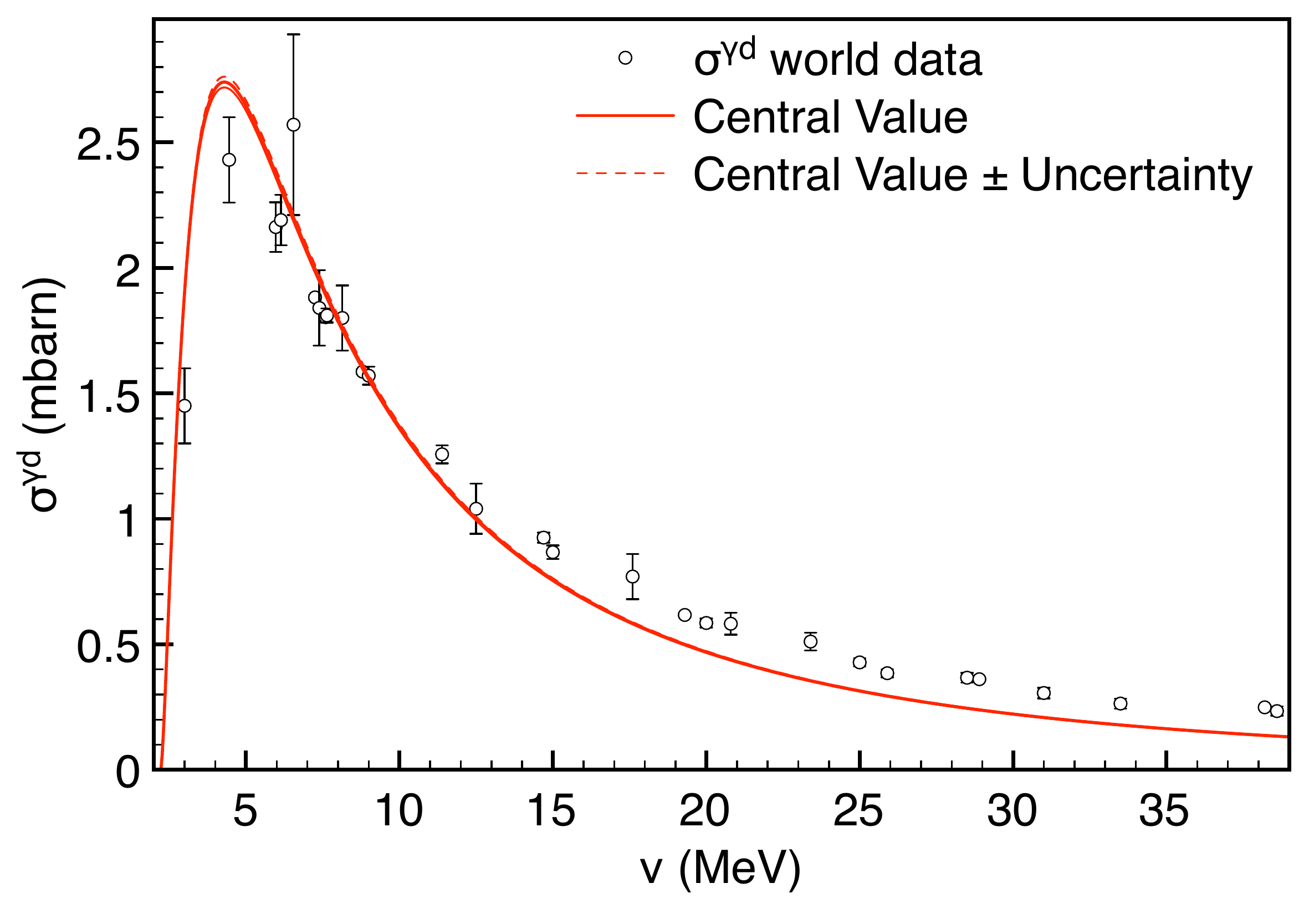}
\caption{(Color online) Same as in Fig. \ref{fig2} vs. deuteron total photoabsorption data from
  Refs. \cite{Bernabei:1986ai,Birenbaum:1985zz}. Older data
  compilation can be found in Ref. \cite{Govaerts:1981xu}}
\label{fig8}
\end{figure}




The parameter $C_\perp$ is obtained from a fit to real photon
data and the value of Baldin sum rule for real photons,
\beqn
\alpha_E^d+\beta_M^d&=&\frac{2\alpha_{em}}{M_d}\int\limits_{\nu_{th}}^{\nu_\pi}\frac{d\nu}{\nu^3}
F^d_1(\nu,0),
\eeqn
where $\alpha_E^d$ and $\beta_M^d$ are the deuteron electric and magnetic polarizabilities, respectively.
There exist
calculations in chiral EFT by Chen et al. \cite{chen} and in
non-relativistic potential models, {\it e.g.} by Friar \cite{friar}, 
that give close results that can be cast in the following form:
$\alpha_E^d=0.633(1)$ fm$^3$, and $\beta_M^d=0.072(5)$ fm$^3$. 
The uncertainty in the value of the polarizabilities was
obtained by averaging over the two calculations. Evaluating Baldin
integral with $F_1^\perp$ (the only piece that does not vanish at the
real photon point) leads to
\beqn
C_\perp=1.28(1).
\eeqn


Adopting these ingredients, the QE contribution of the two-photon exchange (TPE) to Lamb shift in
deuterium can be calculated. 
We evaluated this contribution with the nucleon form factors in 
Kelly's parametrization \cite{kelly} 
and using $S(\nu,Q^2),\,S_\perp(\nu,Q^2)$ from Paris WF \cite{lacomb},
and list the result with the uncertainty in Table \ref{tab2}.  \\

\subsection{Subtraction term}
Following Ref. \cite{Birse:2012eb}, we identify
\beqn
\bar T_1(0,Q^2)&=&T_1^B(0,Q^2)-T_1^{pole}(0,Q^2)\nn\\
&+&\frac{Q^2}{e^2}\beta_M^d(0)F_\beta(Q^2),
\eeqn
where $T^B$ represents the Born contribution, and where in the polarizability term  we explicitly factored out the  $Q^2$-dependence. The polarizability contribution to the $nS$-level is given by 
\beqn
\Delta
E^{\beta}_{n0}=2\alpha\phi^2_{n0}(0)\beta_M^d(0)\int\limits_0^\infty
dQ^2\frac{\gamma_1(\tau_l)}{\sqrt{Q^2}}F_\beta(Q^2),
\eeqn
with $\beta_M^d(0)=0.072(5)$ fm$^3$. 
The $Q^2$-dependent form factor $F_\beta(Q^2)$ is generally not
known. We estimate it by setting $F_\beta(Q^2)=G_C^d(Q^2)/G_C^d(0)$, 
and to estimate the uncertainty we also try
$F_\beta(Q^2)=G_M^d(Q^2)/G_M^d(0)$. 
The average result and uncertainty is quoted in Tab. 2.

Finally, the subtraction function Eq.30 contains the difference between the Born and pole contributions, which results 
from the contact two-photon deuteron interaction (Thomson term). 
The pointlike part of it, $-1/4\pi M_d$ was
already taken into account in atomic calculations, thus we need to
account for
\beqn
&&[T_1^B-T_1^{pole}](0,Q^2)-[T_1^{B,\,point}-T_1^{pole,\,point}](0,Q^2)\nn\\
&&=\frac{1-G_C^2(Q^2)}{4\pi M_d},
\eeqn
thus leading to the shift of an $S$-level
\beqn
\Delta
E_{n0}^{Th}&=&\frac{2\alpha^2}{M_d}\phi^2_{n0}(0)\int_0^\infty \!\!\!dQ^2
\frac{\gamma_1(\tau_l)}{\sqrt{Q^2}}
\frac{1-G_C^2(Q^2)}{Q^2}\label{thomson}
\eeqn
\indent
The result of the numerical evaluation is listed in Table \ref{tab2}.

\begin{table}
  \begin{tabular}{c|c}
\hline
%
%
%
%
$\Delta \bar E^{el}$& -- 0.417(2) \\
\hline
%
$\Delta E^{PWBA}$& -- 1.616(739) \\
\hline
%
$\Delta E^{FSI}$& -- 0.391(44) \\
\hline
%
$\Delta E^{\perp}$& -- 0.322(3) \\
\hline
%
$\Delta E^{hadr}$& -- 0.028(2) \\
\hline
%
$\Delta E^{\beta}$& 0.740(40)\\
\hline
%
$\Delta E^{Th}$& 0.023(1) \\
\hline
\hline
$\Delta E_{total}$& -- 2.011(740)\\
\hline
  \end{tabular}
\caption{TPE corrections to the $2S_{1/2}$ energy level in muonic
deuterium in units of meV.}
\label{tab2}
\end{table}



\section{Discussion of Results and Impact of Further Scattering Experiments}


The total result for the $2P-2S$ Lamb shift obtained from 
the sum of all terms ${\cal O}(\alpha^5)$ due to
two-photon exchange amounts to
\beqn
\Delta E _{2P-2S}&=&2.01(74)\;{\rm meV}.
\eeqn
\indent
The uncertainty of our result comes from three sources: elastic
deuteron form factors, inelastic hadronic excitations and nuclear
(quasi-elastic) contributions. 
The deuteron elastic form factors have been measured over a wide $Q^2$-range with good
precision, and the error associated with different parametrizations of
these data amounts to $2\mu$eV or relative 2\% uncertainty. 
The hadronic part contribution is constrained to a relative 7\%,
however fortunately the contribution itself is rather small, so 
this somewhat large relative uncertainty translates in $2\mu$eV
absolute uncertainty. 

At the moment, for the calculation of the subtraction contribution we rely on the
$Q^2$-dependence for the magnetic polarizability obtained from a
model. A direct calculation of $\bar T_1(0,Q^2)$, for instance in
chiral EFT would help reducing the corresponding uncertainty. 

\begin{figure}
\includegraphics[width=7.5cm]{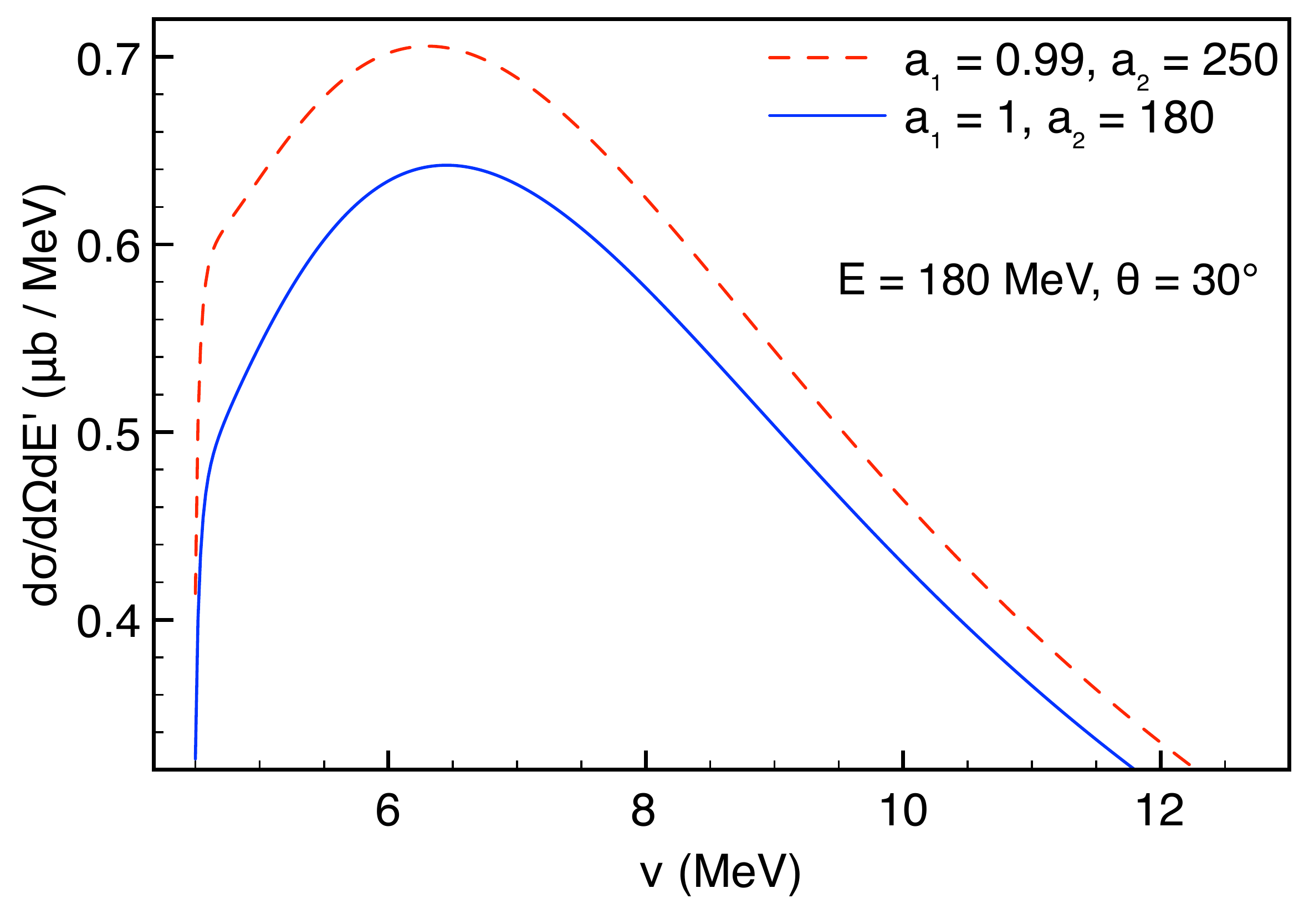}
\includegraphics[width=7.5cm]{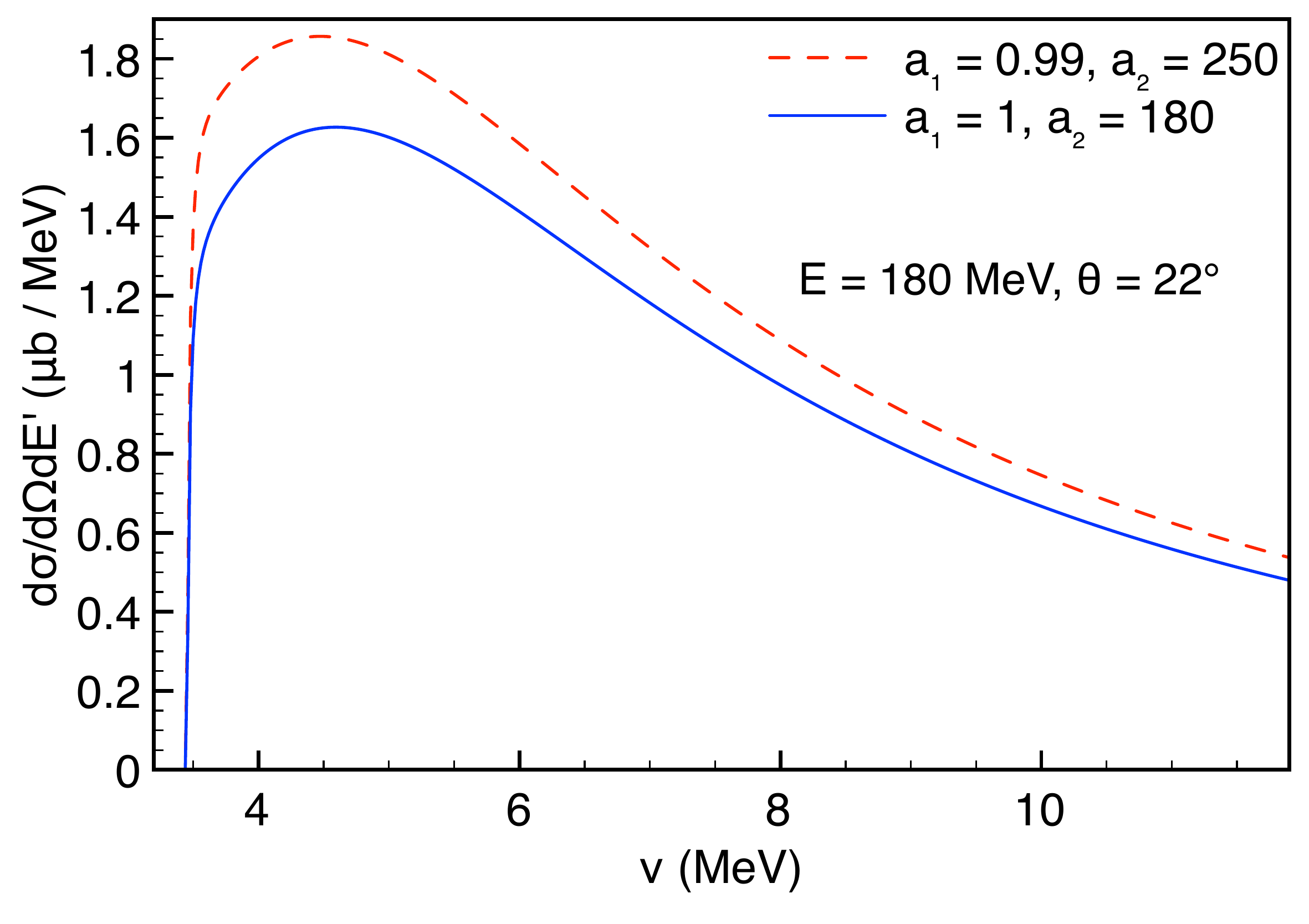}
\includegraphics[width=7.5cm]{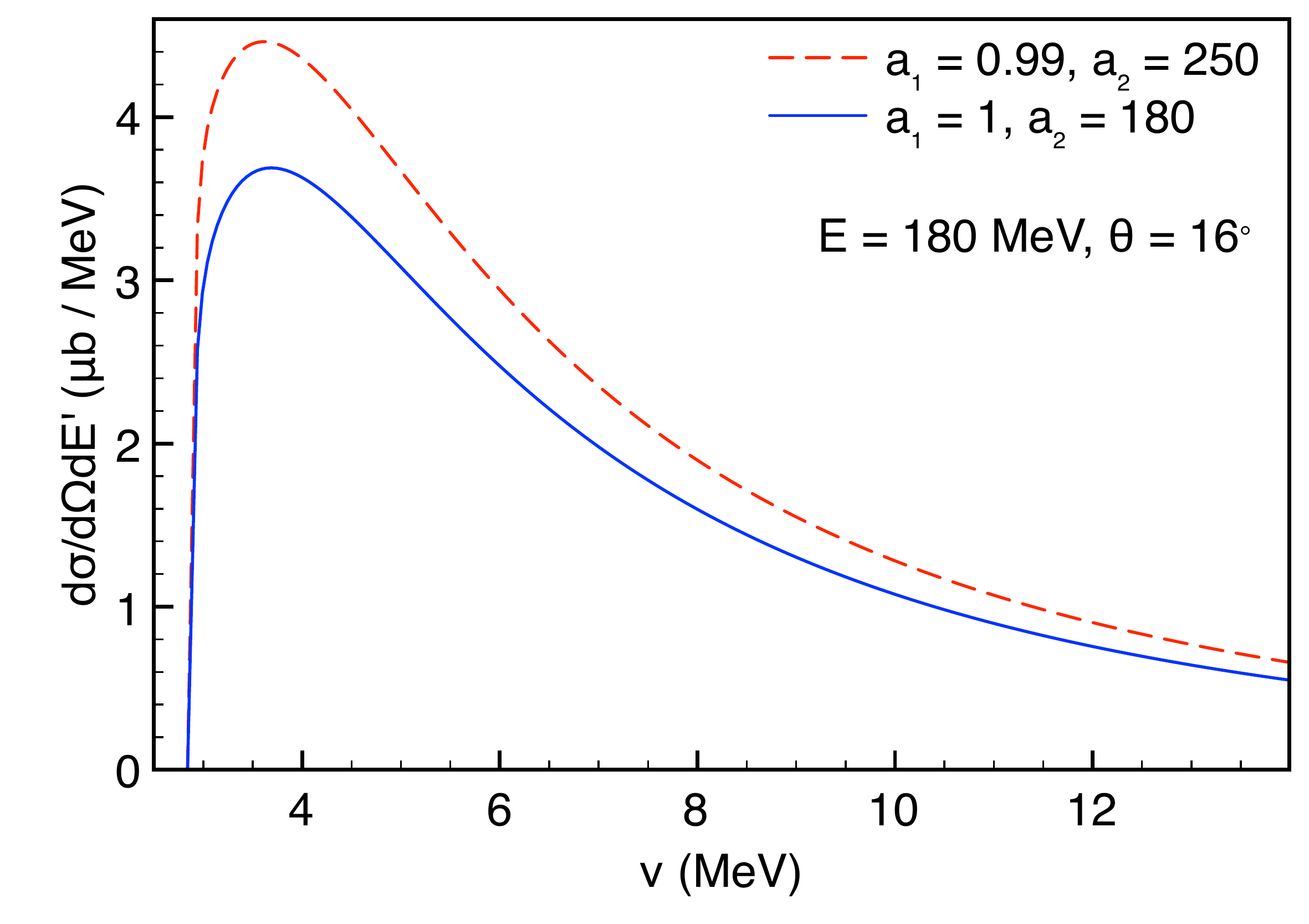}
\includegraphics[width=7.5cm]{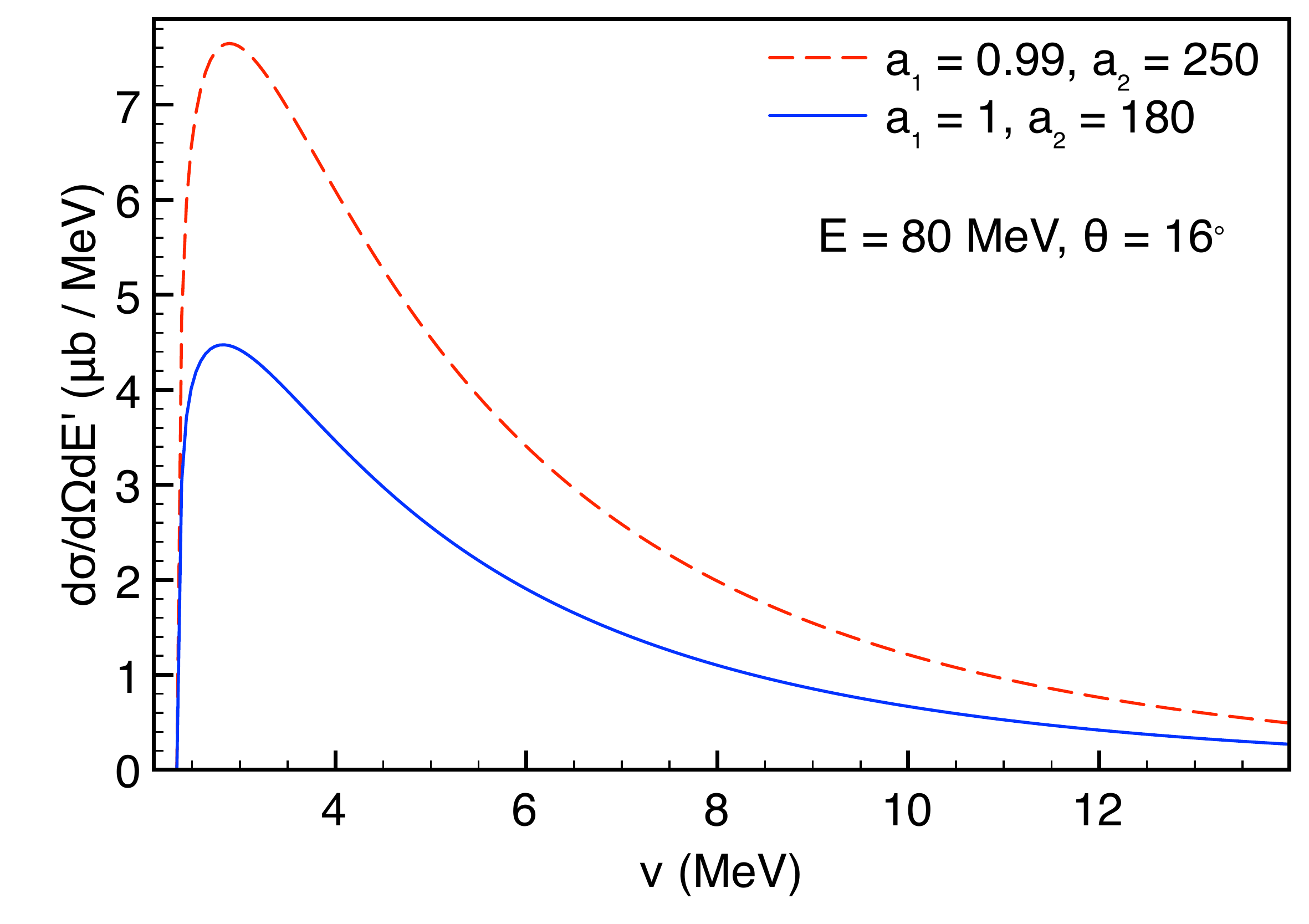}
\caption{(Color online) Sensitivity to the variation of the parameter $a_1$ entering $f_T^{PWBA}$ 
in the range [0.99, 1] 
and $a_2$ entering $f_T^{FSI}$ in the range [180, 250] is shown by the dashed and solid lines, respectively, in the
kinematics relevant for the MAMI A1 apparatus \cite{A1} (three upper
panels), and for MESA at 80 MeV (lower panel).}
\label{sensitivity}
\end{figure}

The largest contribution and the source of the largest
uncertainty is the quasielastic piece, in particular the
$Q^2$-dependence of the inelastic 
structure function $F_2(\nu,Q^2)$ in the range $\nu\leq10$ MeV,
$Q^2\leq0.01$ GeV$^2$ from which the dominant contribution to the Lamb
shift stems. A dedicated measurement at Mainz with the existing A1
apparatus at $E_0=180$ MeV and angles $\theta_{lab}\geq15^\circ$ is
planned \cite{A1}, and it would help somewhat to constrain the
uncertainty with $Q^2\gtrsim2.2\times10^{-3}$ GeV$^2$. 
Going to lower
energies will be possible with the new linear accelerator machine MESA
at Mainz, and we include a few plots demonstrating the sensitivity to 
the parameter $a_1$ in several representative kinematics in Fig. \ref{sensitivity}.

To bring the discussion to a more quantitative level, we list the
projected impact of a $d(e,e')pn$ measurement in several kinematics of
A1 and MESA for the uncertainty of the dispersion calculation of the
Lamb shift in Table \ref{tab3}. For this analysis, we assumed for simplicity that the
uncertainty of the fit will be equal to the precision of the
data.\footnote{If the experimental uncertainty is dominated by the
  systematics this will be a correct estimate. In the opposite case
  the fit to 2\% data will typically return an uncertainty of at most
  1\%.} 
For the kinematics $E_{lab}=80$ MeV, $\theta=16^\circ$ the uncertainty
of the quasielastic contribution is reduced by a factor of 15 and the theory 
uncertainty starts being dominated by that due to the subtraction
constant (estimated to be 40 $\mu$eV). It can be seen that already the
next MAMI A1 runs at the  lowest energy of 180 MeV and the most forward
angle of 16$^\circ$ with a 2\% precision have the potential to reduce the
uncertainty of our dispersion calculation by at least factor of 4. 
The sensitivity to the value of the parameter $a_1$ is further enhanced 
at a lower energy as can be seen in the lower panel of Fig. \ref{sensitivity}. 
Future measurements will allow to test other theoretical frameworks, such as 
potential models and EFT, as well.
\begin{table}
  \begin{tabular}{c|c|c|c}
\hline
$E_{lab},\,\theta_{lab}$ & Exp. precision & 
$\begin{array}{c} \delta(\Delta E_{2S-2P}^{\mu D}) \\ {\rm in}\;\mu {\rm eV} \end{array}$  & 
$\begin{array}{c} \delta(\Delta E_{1S-2S}^{e D}) \\ {\rm in\;kHz} \end{array}$ \\
\hline
\hline
180 MeV, 30$^\circ$ &2\% & 740 & 12 \\
 & 1\% & 370 & 6 \\
\hline
180 MeV, 22$^\circ$ &2\% & 390 & 6.32 \\
 & 1\% & 195 & 3.16 \\
\hline
180 MeV, 16$^\circ$ &2\% & 211 & 3.36 \\
 & 1\% & 110 & 1.68 \\
\hline
80 MeV, 16$^\circ$ &2\% & 67  & 1.08\\
 & 1\% & 48  & 0.78 \\
\hline
 \end{tabular}
\caption{Impact of future measurements of the deuteron
  electrodesintegration at MAMI A1 and MESA (kinematics in the first column and experimental precision in the second column) on the theoretical uncertainty of the TPE contribution to the Lamb shift in muonic deuterium (third column) and the ($1S-2S$) splitting in electronic deuterium (fourth column).}
\label{tab3}
\end{table}

\begin{table}
  \begin{tabular}{c|c|c|c|c|c|c}
\hline
Contribution & This work & \cite{pachuckiD} & \cite{Friar:2013rha} & \cite{Ji:2013oba} & 
\cite{rosenfelder} & \cite{martynenkofaustov} \\
\hline
Elastic & 0.394(2)  & -- & -- & -- & -- & 0.37\\
\hline
Hadronic & 0.028(2) & 0.043 & -- & -- & -- & -- \\
\hline
Nuclear & 1.589(740) & 1.637(16) & -- & -- & 1.5 & -- \\
\hline
\hline
Total & 2.011(740) & 1.680(16) & 1.942 & 1.698 & -- & -- \\
\hline
 \end{tabular}
\caption{Nuclear and nucleon structure-dependent ${\cal{O}}(\alpha^5)$ contributions to the $2P-2S$ Lamb shift in muonic deuterium as calculated by different groups, in units of meV. In case of Refs. \cite{pachuckiD,Friar:2013rha,Ji:2013oba} the separation of the result into ``elastic" and ``nuclear" contributions is not possible, and the sum of the two is quoted. }
\label{tab4}
\end{table}

Our result should be compared to those obtained by other groups:
\cite{pachuckiD,martynenkofaustov,Borie:2012zz,rosenfelder,krutov,friar,Friar:2013rha,Ji:2013oba}. 
Note that \cite{martynenkofaustov,Borie:2012zz,krutov} did not perform a 
complete calculation and take, for instance, the nuclear
polarizability correction from other works. To facilitate the comparison, 
we list our results along with those obtained by other groups in Table \ref{tab4}. To make a sensible comparison possible we reorganized the various contributions listed in Table II as follows: "Elastic" denotes $\Delta\bar E^{el}+\Delta E^{Th}$, and "Nuclear" is sum all nuclear contributions, $\Delta E^{PWBA}+\Delta E^{FSI}+\Delta E^\perp+\Delta E^\beta$.

Ref. \cite{rosenfelder}
quotes 1.500 meV $2P-2S$ correction due to the deuteron nuclear electric
dipole polarizability; in Ref. \cite{pachuckiD} a result of 1.680(16) meV
is obtained by considering the electric polarizability (and various
corrections thereto), elastic and hadronic contributions, and magnetic
polarizability. Ref. \cite{pachuckiD} furthermore obtains the sum of
the proton and neutron intrinsic polarizabilities to the Lamb shift in
muonic deuterium by rescaling the {\it total} Lamb shift for muonic
hydrogen, $\Delta E_{\mu H}=36.9\,\mu$eV obtained in Ref. 
\cite{Carlson:2011zd} with the ratio  $(\mu_r^D/\mu_r^H)^3$ with the
result $\Delta E_{\mu D}^{hadr.}=43(3)\,\mu$eV. 
 This estimate is not correct because the main contribution to $\Delta
E_{\mu H}$ is due to the elastic contribution, and only about a third
of it, $13.5\,\mu$eV comes from polarizabilities. Since proton and
neutron electric polarizabilities are very close,
$\alpha_p\approx\alpha_n$, one should expect that the result for the
deuteron should be roughly equal to their sum, $\Delta E_{\mu D}^{hadr}\sim 2\Delta E_{\mu
  H}=27\,\mu$eV. Indeed, our result (third entry in Table \ref{tab2})
is consistent with this simple estimate, 
$\Delta E_{\mu D}^{hadr}=28(2)\,\mu$eV. This suggests that after
correction the full result of Ref. \cite{pachuckiD} should be 
1.665(16) meV. On the other hand, Ref. \cite{Friar:2013rha} estimates
the Lamb shift in the zero-range approximation to be 1.912 meV 
(1.942 with further corrections), and quotes the result of
Ref. \cite{pachuckiD} in that approximation as 1.899 meV. These numbers
are close to each other, nevertheless, we point out that the
differences are not small, especially compared to the uncertainty of
$16\,\mu$eV claimed in Ref. \cite{pachuckiD}. As mentioned above,
the correct account of the nucleon polarizability corrections alone
shifts the result of Ref. \cite{pachuckiD} by $15\,\mu$eV that exhausts the claimed
precision of the calculation. In Ref.~\cite{Ji:2013oba} the calculation of the 
polarizability correction is reexamined and higher-order relativistic corrections 
from longitudinal and transverse two-photon exchanges were included, 
leading to an additional contribution of $18\,\mu$eV.
 

\section{Electronic Hydrogen}
To complete the discussion, we assess 
the nuclear polarizability correction for 
the $nS$-levels in the usual (electronic) deuterium, too. In
particular, the isotopic shift measurement of $1S-2S$ splitting of
Ref. \cite{Parthey:2010} relies on the theoretical estimate 
according to Ref. \cite{FriarPayne}, 
\beqn
\Delta E_{2S-1S}^{e-D}&=&19.04(7)\,{\rm kHz},
\eeqn
where the polarizability correction of 18.58(7) kHz and the elastic
contribution of 0.46 kHz were added together. 
Ref. \cite{milshtein}
gives a somewhat different result,
\beqn
\Delta E_{2S-1S}^{e-D}&=&19.25\,{\rm kHz},
\eeqn
with the Coulomb contribution 17.24 kHz, the magnetic contribution
2.28 kHz and the magnetic polarizability correction -0.27 kHz. 

Our evaluation for the $1S-2S$
splitting in deuterium is
\beqn
\Delta E_{2S-1S}^{e-D}&=&28.8\pm12.0\,{\rm kHz}, 
\eeqn
that is the sum of the elastic (0.53(1) kHz), inelastic (33.4(12.0) kHz)
and subtraction (-4.60(3) kHz) contributions. The uncertainty is about a half 
of the full result. Since for the electronic deuterium the
integrals over structure functions are even more strongly weighted at low values of
$Q^2$ where no experimental information is available, the large uncertainty does not come unexpectedly. We show in Table \ref{tab3} (fourth column) how future electron-deuteron scattering measurements can help improving on this estimate. 

Note that this uncertainty estimate exceeds the one in Eq. (34) by two orders of magnitude.  However, the main uncertainty in the isotope shift given in~\cite{Parthey:2010} is actually due to uncertainties in other theoretical corrections, largely caused by uncertainties in parameters such as particle masses.  The total radius-related energy uncertainty in~\cite{Parthey:2010} is $0.89$ kHz.  The uncertainty from the dispersive polarizability calculation is still an order of magnitude larger; using it would change the radius difference result to
\beqn
r_E^2(d) - r_E^2(p) = 3.8274(88) \, {\rm fm}^2,
\eeqn
increasing the uncertainty by a factor of $\sim10$ as compared to Eq. (1).
Using the CODATA value for the proton charge radius $r_E(p)=0.8775(51)$ fm leads us to a new extracted value of the deuteron radius,
\beqn
r_E(d)=2.1442(29) \, {\rm fm},
\eeqn
that should be compared to the previous extraction \cite{Mohr:2012tt}, $r_E(d)=2.1424(21)$ fm. 
Thus, in the electron case the increase in the polarizability uncertainty for the isotope shift makes it comparable to the existing uncertainty in the proton radius-squared.  Using it merely increases the uncertainty in the inferred deuteron radius by a factor of $\sqrt2$.



\section{Conclusion}


We conclude that in the case of deuterium, 
model independence that is the main objective of our
approach comes at a high price. Scattering data do not constrain the
behavior of structure functions, especially the longitudinal one, at
low values of the momentum transfer. Microscopical nuclear
calculations do a much better job in terms of intrinsic precision that
typically is of order of fractions of a per cent. However, in absence of data 
this claimed precision is not warranted, and once new low-$Q^2$
electrodisintegration data will be available they will
serve as a useful cross check for nuclear calculations, as well.

\begin{acknowledgments}
We are grateful to M. Distler, Z.-E. Meziani, V. Pascalutsa, S. Karshenboim, D.R. Phillips, J. Yang, H. Griesshammer, K. Pachucki, J.L. Friar and S. Bacca for useful discussions.
The work of M.G. and M.V. was supported by the Deutsche Forschungsgemeinshaft DFG through the Collaborative Research Center ``The Low-Energy Frontier of the Standard Model" (SFB 1044) and the Cluster of Excellence ``Precision Physics, Fundamental Interactions and Structure of Matter" (PRISMA). C.E.C. acknowledges support by
the U.S. National Science Foundation under Grant PHY-1205905.
\end{acknowledgments}

\end{document}